\documentclass[aps,prl,reprint,groupedaddress]{revtex4-1}
\usepackage{graphicx}
\usepackage{amsmath,amssymb}
\usepackage{amsfonts}
\usepackage{mathtools}
\usepackage[utf8]{inputenc}
\usepackage[colorlinks=true,linkcolor=blue,citecolor=blue,filecolor=cyan,urlcolor=blue,breaklinks=true]{hyperref}
\usepackage{subfigure}
\usepackage{psfrag}
\usepackage{color}
\usepackage{microtype}
\DisableLigatures[f]{encoding = *, family = *}
\newcommand{\ip}[2]{\langle #1,#2\rangle_\pi}
\newcommand{\HH}{\mathcal H_0}
\newcommand{\EE}{\mathcal E}
\newcommand{\Var}{\operatorname{Var}}
\newcommand{\Rea}{\operatorname{Re}}

\newcommand{\ii}{\mathrm i}
\newcommand{\pos}[1]{\left[#1\right]_+}
\newcommand{\heading}[1]{\emph{#1}.---}
\definecolor{boundblue}{RGB}{0,50,200}
\definecolor{boundred}{RGB}{177,50,48}
\definecolor{boundgreen}{RGB}{31,119,91}

\begin{document}
\title{Fundamental Bounds on the Oscillation Efficiency of Markov Networks}
\author{Chen Jia$^{1,2}$}\email{Correspondence: chenjia@hit.edu.cn}
\affiliation{$^1$ School of Mathematics, Harbin Institute of Technology, Harbin 150001, China.}
\affiliation{$^2$ Beijing Research Institute, Harbin Institute of Technology, Beijing 100144, China.}

\begin{abstract}
Stochastic oscillations are commonly identified through peaks in their power spectra, yet the fundamental limits on their strength and energetic cost remain unclear. Here we establish universal dynamical and thermodynamic bounds on oscillation efficiency in Markov networks, showing how finite state spaces constrain spectral peak enhancement. In particular, we demonstrate that networks maximizing modal coherence need not maximize oscillation efficiency. We further derive dissipation bounds that connect oscillation efficiency to the relaxation speed and stationary probabilities, revealing an unavoidable energetic cost of maintaining pronounced oscillations at a specified relaxation timescale. These bounds remain strictly positive in regimes where existing explicit spectral dissipation estimates vanish, with stronger constraints for normal networks and uniform rings. Our results distinguish spectral oscillatory strength from modal coherence and reveal how network dynamics and thermodynamic dissipation jointly limit stochastic oscillations.
\end{abstract}

\maketitle

\heading{Introduction}
Sustained oscillations are common in nature and coordinate biological rhythms across a broad range of scales. Some of the best-understood biological oscillators include p53 responses \cite{lahav2004dynamics}, NF-kB responses \cite{nelson2004oscillations}, cell cycle \cite{li2004yeast}, circadian rhythm \cite{lee2000interconnected}, and various synthetic oscillators \cite{elowitz2000synthetic, stricker2008fast}. However, there are growing observations that biological oscillations typically display stochastic fluctuations, which can play a vital role in enhancing or weakening the robustness of these oscillations \cite{mckane2005predator, bratsun2005delay, thomas2013signatures, gupta2022frequency, jia2024exact}.

Markov networks are widely used to describe molecular machines, biochemical reactions, and regulatory circuits, where intrinsic fluctuations give rise to pronounced cell-to-cell variability \cite{seifert2012stochastic}. The power spectrum provides a central tool for studying stochastic oscillations in Markov networks \cite{mckane2005predator, bratsun2005delay, thomas2013signatures, gupta2022frequency, jia2024exact}. A nonzero-frequency peak of the power spectrum identifies a dominant frequency in a noisy signal, and its height relative to the zero-frequency background measures the robustness of oscillations. For equilibrium systems obeying detailed balance, the power spectrum is a positive sum of Lorentzians and always decreases monotonically; hence a nonmonotonic spectrum is a signature of broken detailed balance and positive entropy production \cite{qian2000pumped, wu2025parameter}.

A natural and crucial question arises: what is the energetic cost of stochastic oscillation characterized by a nonmonotonic power spectrum? Recently, fundamental thermodynamic bounds have illuminated the dissipation costs and physical limits of the precision, speed, and persistence of nonequilibrium functions from diverse perspectives, including thermodynamic uncertainty relations, speed limits, and inference from partially observed transitions \cite{barato2015thermodynamic, gingrich2016dissipation, shiraishi2018speed, skinner2021estimating, van2022thermodynamic, harunari2022learn, van2023time, van2023thermodynamic}. Advances addressing temporal correlations, oscillatory dynamics, and spectral properties have further revealed how dissipation constrains the temporal organization of nonequilibrium systems \cite{barato2017coherence, oberreiter2022universal, ohga2023thermodynamic, dechant2023thermodynamic, kolchinsky2024thermodynamic, vo2025inverse}.

For stochastic oscillations, previous work has linked cycle coherence to thermodynamic driving and network topology \cite{barato2017coherence,oberreiter2022universal,ohga2023thermodynamic}, and has developed spectral estimates of dissipation and finite-frequency fluctuation-response inequalities \cite{dechant2023arxiv,dechant2026finite}. However, the maximum achievable oscillation efficiency and its energetic cost remain unclear; existing explicit spectral estimates can even vanish for nonmonotonic power spectra \cite{dechant2023arxiv}. In this study, we derive universal dynamical and thermodynamic bounds on oscillation efficiency in finite Markov networks, revealing the tradeoff between oscillation robustness, relaxation speed, and dissipation. These results establish physical limits on the strength and cost of stochastic oscillations.

\begin{figure}[htb!]
\centering\includegraphics[width=0.48\textwidth]{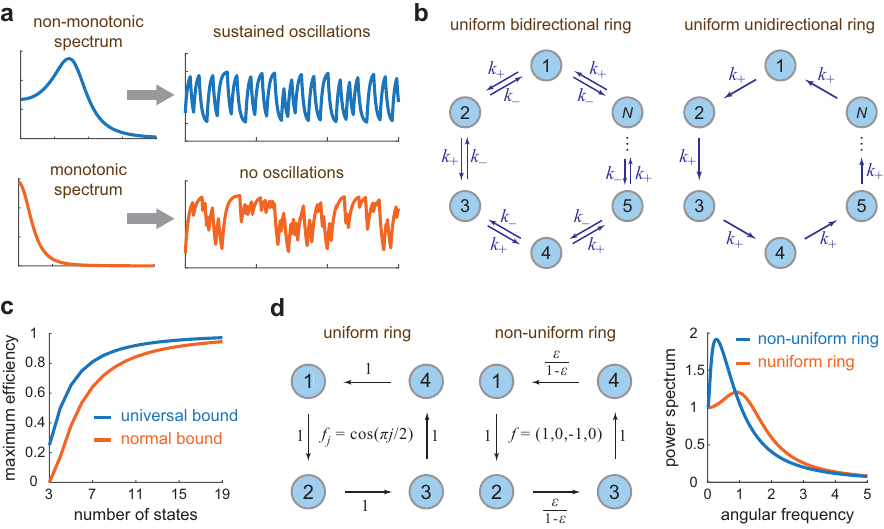}
\caption{\textbf{Dynamic bounds on the oscillation efficiency in Markov networks.}
\textbf{(a)} Non-monotonic power spectrum serves as a characteristic signature of sustained stochastic oscillations.
\textbf{(b)} Uniform bidirectional and unidirectional rings.
\textbf{(c)} Comparison of the universal bound in Eq.~\eqref{eq:dynamic} and the normal bound in Eq.~\eqref{eq:normaldynamic}.
\textbf{(d)} For four-state systems, non-uniform rings may produce a larger oscillation efficiency than uniform rings. Here the power spectrum is normalized so that $G_f(0)=1$. For the non-uniform ring, we choose $\epsilon = 0.05$.}
\label{figure1}
\end{figure}

\heading{Model and oscillation efficiency}
Consider an $N$-state molecular system modeled by a continuous-time Markov chain (also called a Markov network \cite{van2022thermodynamic}) $(X_t)_{t\ge 0}$ with states $i = 1,\cdots,N$ and generator matrix $Q = (q_{ij})$, where $q_{ij}$ denotes the transition rate from state $i$ to $j$ whenever $i\neq j$ and $q_{ii} = -\sum_{j\neq i}q_{ij}$. We assume that the system is stationary and irreducible. The unique stationary distribution $\pi=(\pi_1,\ldots,\pi_N)$ obeys $\pi Q=0$ and $\pi_i>0$ for all $i$. For any real-valued, nonconstant state observable $f$, recall that the autocorrelation function of the system, $C_f(t) = \mathrm{Cov}(f(X_0),f(X_t))$, is defined to be the stationary covariance of $f(X_0)$ and $f(X_t)$, and the power spectrum $G_f(\omega)=\int_{-\infty}^{\infty}C_f(t)e^{-\mathrm{i}\omega t}\,\mathrm{d}t$ is defined as the Fourier transform of $C_f(t)$, where $\omega\geq 0$ denotes the angular frequency. For convenience, in what follows, we subtract the stationary mean from $f$ and consider the centered observable, so $\sum_i\pi_i f_i=0$.

A nonmonotonic $G_f(\omega)$ is a characteristic signature of stochastic oscillations, with the peak location defining the dominant frequency (Fig.~\ref{figure1}(a)). Writing $G_f^{\max}=\max_{\omega\ge0}G_f(\omega)$, we define the \emph{oscillation efficiency} as \cite{jia2024exact}
\begin{equation}\label{eq:eta}
\eta_f  = \frac{G_f^{\max}-G_f(0)}{G_f^{\max}} = 1-\frac{G_f(0)}{G_f^{\max}},
\end{equation}
which lies between $0$ and $1$. Clearly, a large $\eta_f$ indicates that the spectral power at the dominant frequency is much greater than that at zero frequency, signifying strong oscillatory behavior. If $G_f(\omega)$ decreases monotonically, then $G_f^{\max}=G_f(0)$ and $\eta_f=0$. Hence $\eta_f$ provides an effective measure of the robustness of oscillations.

\heading{Universal dynamical bounds}
A key question is whether the oscillation efficiency of Markov networks has a fundamental limit. Let $\eta_N^*:=\sup_{Q,f}\eta_f$ denote the maximum oscillation efficiency for all $N$-state Markov networks across all state observables. For any $N\geq 3$, we find that $\eta_N^*$ has the following universal upper bound:
\begin{equation}\label{eq:dynamic}
\eta_N^* = \sup_{Q,f}\eta_f \le\cos^2\frac\pi N.
\end{equation}

We now give the proof of this fact. In Ref.~\cite{gupta2022frequency}, it was shown that the power spectrum can be represented as
\begin{equation}\label{eq:psd}
G_f(\omega) = 2\Rea\ip{f}{(A+\ii\omega I)^{-1}f},
\end{equation}
where $\ip{u}{v}=\sum_i\pi_i\overline{u}_iv_i$ denotes the inner product with respect to the stationary distribution $\pi$, and $A=-Q$ is an operator restricted to the space $\HH=\{u:\sum_i\pi_i u_i=0\}$ of all centered complex-valued state observables. All inverses below are taken on $\HH$. To proceed, we decompose $A$ as the sum of its self-adjoint and skew-adjoint parts, i.e.
\begin{equation}\label{eq:operators}
A=H+K,\quad
H=\frac{A+A^*}{2},\quad
K=\frac{A-A^*}{2},
\end{equation}
where $A^*$ denotes the adjoint operator with respect to the inner product $\langle\cdot,\cdot\rangle_{\pi}$, and let $B=H^{-1/2}KH^{-1/2}$ be a dimensionless operator which measures irreversible circulation relative to reversible relaxation. It is easy to see that $H$ is self-adjoint and positive definite on $\HH$, and
\begin{equation}
A+\ii\omega I = H^{1/2}(I+B+\ii\omega H^{-1})H^{1/2}.
\end{equation}
Let $h=H^{-1/2}f$ and define
\begin{equation}
T_\omega=B+\ii\omega H^{-1},\qquad z_\omega=(I+T_\omega)^{-1}h.\
\end{equation}
Eq.~\eqref{eq:psd} then reads $G_f(\omega)/2 = \Rea\ip{h}{z_\omega}$. Since $T_\omega$ is skew-adjoint, $\Rea\ip{z_\omega}{T_\omega z_\omega}=0$. Using $h=z_\omega+T_\omega z_\omega$, we obtain the following two identities:
\begin{equation}
\frac{G_f(\omega)}{2}=\|z_\omega\|_\pi^2,\quad
\|h\|_\pi^2=\|z_\omega\|_\pi^2+\|T_\omega z_\omega\|_\pi^2.
\end{equation}
They immediately imply
\begin{equation}\label{eq:peakB}
G_f(\omega)=2\|z_\omega\|_\pi^2
\le 2\|h\|_\pi^2.
\end{equation}
At zero frequency, $T_0=B$. The same identities give
\begin{equation}
\|h\|_\pi^2
=\|z_0\|_\pi^2+\|Bz_0\|_\pi^2
\le(1+\|B\|^2)\frac{G_f(0)}{2},
\end{equation}
where $\|B\| = \sup_{u\neq 0}(\|Bu\|_{\pi}/\|u\|_{\pi})$ denotes the operator norm of $B$. This clearly shows that
\begin{equation}\label{eq:zeroB}
G_f(0)\ge\frac{2\|h\|_\pi^2}{1+\|B\|^2}.
\end{equation}
Moreover, in the End Matter, we use cycle decomposition techniques of Markov networks to prove that
\begin{equation}\label{eq:sectorN}
\|B\|\le\cot\frac{\pi}{N}.
\end{equation}
Combining Eqs.~\eqref{eq:peakB}, \eqref{eq:zeroB}, and \eqref{eq:sectorN}, we finally obtain
\begin{equation}\label{eq:etaB}
\eta_f
=1-\frac{G_f(0)}{G_f^{\max}}
\le\frac{\|B\|^2}{1+\|B\|^2}
\le\cos^2\frac{\pi}{N}.
\end{equation}

For $3$-state and $4$-state networks, we prove in the End Matter that the upper bound in Eq.~\eqref{eq:dynamic} is sharp, i.e.
\begin{equation}\label{eq:smallN}
\eta_3^*=1/4,\qquad \eta_4^*=1/2.
\end{equation}
For $N\ge5$, the upper bound may still be sharp, but now we are unable to give a rigorous proof.

\heading{Stronger bounds for normal networks}
There is a stronger upper bound on oscillation efficiency if we further require the generator matrix $Q$ to be a normal operator, namely $QQ^*=Q^*Q$. In \cite{supp}, we prove that
\begin{equation}\label{eq:normaldynamic}
\max_{Q\ {\rm normal},f}\eta_f
=\left(2\cos\frac\pi N-1\right)^2.
\end{equation}
An important subclass of normal networks is the \emph{uniform bidirectional ring} with forward and backward transition rates $k_+$ and $k_-$, i.e. $q_{i,i+1}=k_+$ and $q_{i,i-1}=k_-$, where indices are understood to be modulo $N$ and $k_+\ge k_->0$ are constants (Fig.~\ref{figure1}(b)). For such systems, the stationary distribution is uniform, i.e. $\pi_i=1/N$, and the generator matrix $Q$ is circulant and hence is normal. Uniform rings provide idealized descriptions of many cyclic biological processes, such as enzyme turnover \cite{min2005nonequilibrium}, rotary ATPase dynamics \cite{yasuda1998f1}, and ion transport through cell membrane pores \cite{schnakenberg1976network}.

Actually, the upper bound in Eq.~\eqref{eq:normaldynamic} is sharp. It is attained when the Markov network is a \emph{uniform unidirectional ring} (Fig.~\ref{figure1}(b)), a special case of the uniform bidirectional ring by setting $k_+>0$ and $k_-=0$, together with the observable $f_i=\cos(2\pi i/N)$. Therefore, the uniform unidirectional ring realizes the maximum efficiency among all $N$-state normal networks. For $N=3$, the bound in Eq.~\eqref{eq:normaldynamic} reduces to zero, so every normal three-state network has zero oscillation efficiency.

Fig.~\ref{figure1}(c) compares the universal and normal dynamical bounds in Eqs.~\eqref{eq:dynamic} and \eqref{eq:normaldynamic} as $N$ varies. The distinction between them is counterintuitive. A uniform unidirectional ring is often regarded as an optimal oscillator because it maximizes oscillation coherence, measured by the ratio of oscillation frequency to the corresponding decay rate, at fixed state number \cite{barato2017coherence}. Yet it does not maximize the oscillation efficiency (Fig.~\ref{figure1}(d)): in the End Matter, we show that a non-normal three-state network can have positive oscillation efficiency even when all eigenvalues are real; moreover, a non-normal four-state network can approach an efficiency of $\eta_4^* = 1/2$, much higher than the maximum $(2\cos(\pi/4)-1)^2\approx 0.17$ for normal networks.

\heading{Universal thermodynamic bound}
While the upper bounds on oscillation efficiency have been obtained above, attaining these bounds may require infinitely large energy dissipation (the entropy production rate of a uniform unidirectional ring is infinite). This raises a crucial question: does there exist a thermodynamic bound on oscillation efficiency for a given level of dissipation?

For each pair of states $i$ and $j$, let $w_{ij} = \pi_iq_{ij}$ denote the probability flow from $i$ to $j$. Moreover, we define
\begin{equation*}
J_{ij}=w_{ij}-w_{ji},\quad \tau_{ij}=w_{ij}+w_{ji},\quad
\sigma_{ij}=J_{ij}\log\frac{w_{ij}}{w_{ji}}.
\end{equation*}
In units with $k_B=1$, recall that the entropy production rate of a Markov network is \cite{seifert2012stochastic}
\begin{equation}\label{eq:sigma}
\sigma = \sum_{i<j}\sigma_{ij} = \sum_{i<j}J_{ij}\log\frac{w_{ij}}{w_{ji}},
\end{equation}
which quantifies the energetic cost of nonequilibrium operation. We now impose \emph{thermodynamic consistency}, namely $q_{ij}>0\Longleftrightarrow q_{ji}>0$ \cite{van2022thermodynamic}. This guarantees that the system has a finite entropy production rate. Sums over edges include only pairs with $\tau_{ij}>0$; absent edges
contribute zero. Our universal thermodynamic bound is
\begin{equation}\label{eq:thermaleta}
\eta_f\le\frac{(1-2\pi_{\min})\sigma}{4\pi_{\min}\lambda_s+(1-2\pi_{\min})\sigma},
\end{equation}
where $\pi_{\min} = \min_i\pi_i>0$ is the minimum stationary probability across all states and $\lambda_s=\lambda_{\min}(H|_{\HH})>0$ is the spectral gap of the operator $H = (A+A^*)/2$, which characterizes the relaxation rate of the system towards the steady state. Since $\pi_{\min}\le1/N$, the coefficient $1-2\pi_{\min}$ is positive for any $N\geq 3$. This bound reveals the maximum efficiency attainable for a given level of dissipation.

Note that Eq.~\eqref{eq:thermaleta} can be rewritten as
\begin{equation}\label{eq:universalthermal}
\sigma\ge\frac{4\pi_{\min}}{1-2\pi_{\min}}\lambda_s
\frac{\eta_f}{1-\eta_f}.
\end{equation}
This reveals the tradeoff between oscillation robustness, relaxation speed, and energy dissipation --- achieving high oscillation efficiency $\eta_f$ and rapid relaxation rate $\lambda_s$ requires greater energetic cost. The appearance of $\lambda_s$ in the above inequality is natural. Multiplying all transition rates of the system by $c>0$ leaves both $\pi$ and $\eta_f$ unchanged, but multiplies both $\sigma$ and $\lambda_s$ by $c$. Slowing an oscillator thus reduces its entropy production rate without changing its oscillation efficiency. Hence a positive lower bound for $\sigma$ cannot follow from a dimensionless oscillation efficiency $\eta_f$ alone. Eq.~\eqref{eq:universalthermal} instead bounds the entropy produced per relaxation time, $\sigma/\lambda_s$.

We next give the proof of Eq.~\eqref{eq:thermaleta}. The elementary inequality
$(x-y)\log(x/y)\ge2(x-y)^2/(x+y)$ gives
\begin{equation}
\sigma_{ij}\ge 2J_{ij}^2/\tau_{ij}.\label{eq:edgeentropy}
\end{equation}
For any real $u\in\HH$, let $\EE(u)=\ip{u}{Hu}$ be the energy functional and set $m_{ij}(u)=(u_i+u_j)/2$. Since $H$ is self-adjoint in $L^2(\pi)$, i.e. $H = H^*$, and since each row of $H$ sums to zero, it is easy to see that
\begin{equation}
\EE(v)=\frac12\sum_{i<j}\tau_{ij}(v_j-v_i)^2.
\end{equation}
Moreover, stationary current conservation yields
\begin{equation}\label{eq:currentform}
\begin{split}
\ip{u}{Kv}
&=-\frac12\sum_{i<j}J_{ij}(u_iv_j-u_jv_i)\\
&=-\sum_{i<j}J_{ij}m_{ij}(u)(v_j-v_i).
\end{split}
\end{equation}
The Cauchy-Schwarz inequality and Eq.~\eqref{eq:edgeentropy} then imply
\begin{equation}\label{eq:bilinearthermal}
|\ip{u}{Kv}|^2\le\EE(v)
\sum_{i<j}\sigma_{ij}m_{ij}(u)^2.
\end{equation}
If $\|B\|=0$, then the system is in detailed balance. In this case, Eq.~\eqref{eq:thermaleta} holds trivially since $\sigma = \eta_f = 0$. On the other hand, if $\|B\|\neq 0$, then the real skew-adjoint operator $B$ has a two-dimensional invariant plane with real orthonormal vectors $x,y\in\HH$ satisfying $Bx=\|B\|y$ and $By=-\|B\|x$. Setting $u=H^{-1/2}x$ and $v=H^{-1/2}y$, we obtain
\begin{equation*}
\EE(u)=\EE(v)=1,\;\;
\ip{u}{Hv}=0,\;\;
|\ip{u}{Kv}|=\|B\|.
\end{equation*}
The same identities also hold for
\begin{equation}
u_t=u\cos t+v\sin t,\qquad v_t=-u\sin t+v\cos t.
\end{equation}
Applying Eq.~\eqref{eq:bilinearthermal} to this rotated pair $(u_t,v_t)$ and averaging over $t\in[0,2\pi]$ gives
\begin{equation}\label{eq:rotation}
\|B\|^2\le\frac12\sum_{i<j}\sigma_{ij}
\bigl[m_{ij}(u)^2+m_{ij}(v)^2\bigr].
\end{equation}
To proceed, we define
\begin{equation}
z^{ij}_\ell=\frac{\mathbf1_{\{\ell=i\}}}{2\pi_i}
+\frac{\mathbf1_{\{\ell=j\}}}{2\pi_j}-1,
\end{equation}
where $\mathbf1_{\{\ell=i\}}$ is the indicator function which takes the value of $1$ when $\ell=i$ and the value of $0$ otherwise. It is easy to verify that $z^{ij}\in\HH$ and $m_{ij}(u)=\ip{z^{ij}}{u}$. Note that $x = H^{1/2}u$ and $y = H^{1/2}v$ are orthonormal. Bessel's inequality then implies
\begin{equation}\label{eq:bessel}
\begin{split}
&m_{ij}(u)^2+m_{ij}(v)^2
\le\ip{z^{ij}}{H^{-1}z^{ij}}
\le\frac{\|z^{ij}\|_{\pi}^2}{\lambda_s}\\
&\quad=\frac1{\lambda_s}\left(\frac1{4\pi_i}+\frac1{4\pi_j}-1\right)
\le\frac{1-2\pi_{\min}}{2\pi_{\min}\lambda_s}.
\end{split}
\end{equation}
Substituting this into Eq.~\eqref{eq:rotation} establishes
\begin{equation}\label{eq:Bthermal}
\|B\|^2\le\frac{1-2\pi_{\min}}{4\pi_{\min}}\frac{\sigma}{\lambda_s}.
\end{equation}
Eqs.~\eqref{eq:peakB} and \eqref{eq:zeroB} imply $\eta_f\le\|B\|^2/(1+\|B\|^2)$. This inequality, together with Eq.~\eqref{eq:Bthermal}, finally yields Eq.~\eqref{eq:thermaleta}.

\heading{Stronger bounds for normal networks and uniform rings}
The thermodynamic bound can be further strengthened if we assume the generator matrix $Q$ to be normal. In \cite{supp}, we prove that for any normal network, the entropy production rate has the lower bound
\begin{equation}\label{eq:normalthermal}
\sigma\ge\frac{4\pi_{\min}}{1-2\pi_{\min}}\lambda_s\Phi(\eta_f),
\end{equation}
where
\begin{equation}\label{eq:Phi}
\Phi(\eta_f)=\frac{(1+\sqrt\eta_f)^2}{(1-\sqrt\eta_f)(3+\sqrt\eta_f)}\mathbf1_{\{\eta_f>0\}}.
\end{equation}
The function $\Phi(\eta_f)$ strictly exceeds $\eta_f/(1-\eta_f)$ on $(0,1)$, and hence Eq.~\eqref{eq:normalthermal} sharpens Eq.~\eqref{eq:universalthermal} under the additional normality assumption.

We next focus on the special case of uniform bidirectional rings (Fig.~\ref{figure1}(b)). Since such rings are normal networks, Eq.~\eqref{eq:normalthermal} provides a lower bound on $\sigma$. Since the stationary distribution is uniform for such rings, we have $\pi_{\min} = 1/N$ and hence the prefactor in Eq.~\eqref{eq:normalthermal} is given by $4\pi_{\min}/(1-2\pi_{\min}) = 4/(N-2)$. In \cite{supp}, we show that this lower bound can be further strengthened to
\begin{equation}\label{eq:ringquadratic}
\sigma\ge\frac{\lambda_s}{\cos^2(\pi/N)}\Phi(\eta_f).
\end{equation}
For $N = 3,4$, we have $1/\cos^2(\pi/N) = 4/(N-2)$ and hence Eqs.~\eqref{eq:normalthermal} and \eqref{eq:ringquadratic} are equivalent. For $N\ge 5$, however, $1/\cos^2(\pi/N)$ strictly exceeds $4/(N-2)$ and hence the ring-specific bound is indeed stronger than the bound for normal networks.

\heading{Comparison with spectral dissipation estimates}
In a previous study \cite{dechant2023arxiv}, Dechant derived lower bounds on the entropy production rate $\sigma$ in terms of the power spectrum $G_f(\omega)$. Specifically, let $V_f=\Var_\pi(f)$ be the stationary variance of the observable, let $\Delta f=\max_i f_i-\min_i f_i$ be the range of the observable, and let $D_f=\lim_{\omega\to\infty}\omega^2G_f(\omega)$ be the high-frequency decay coefficient of the power spectrum. Two explicit spectral estimates of dissipation derived from Ref.~\cite{dechant2023arxiv} are given by
\begin{equation}\label{eq:dechant}
\begin{gathered}
\sigma\ge L_D^{(1)}
:=\frac2{(\Delta f)^2}
\sup_{\omega\ge0}\pos{\omega^2G_f(\omega)-D_f},\\
\sigma\ge L_D^{(2)}
:=\frac2{(\Delta f)^2}
\pos{\frac{4V_f^2}{G_f(0)}-D_f},
\end{gathered}
\end{equation}
with $\pos{x} = \max\{0,x\}$. Here the first bound detects an overshoot of the weighted spectrum $\omega^2G_f(\omega)$ above its high-frequency limit; the second detects sufficiently suppressed zero-frequency fluctuations. However, both estimates may yield a zero lower bound for stochastic oscillatory systems.

To illustrate this limitation, we consider an idealized three-site phosphorylation model inspired by KaiC in the cyanobacterial circadian clock \cite{rust2007ordered}. In addition to its canonical phosphorylation sites S431 and T432, phosphorylation at an alternative site, S320, has been observed in a KaiC phosphomimetic mutant \cite{pattanayek2011combined}. Motivated by this observation, we model two canonical sites by a four-state phosphorylation cycle $X_t\in\{U,T,ST,S\}$ with forward and backward transition rates $k_+$ and $k_-$ \cite{rust2007ordered}, and model an additional site by an independent two-state phosphorylation process $Y_t\in\{0,1\}$ (Fig.~2(a)). For analytical tractability, we assume that the additional site switches between its unphosphorylated and phosphorylated states at a symmetric rate $r$. The generator matrix of the coupled Markov network takes the form
\begin{equation}
Q=Q_X\otimes I+I\otimes Q_Y,
\end{equation}
where $Q_X$ and $Q_Y$ are the generator matrices of $X_t$ and $Y_t$, respectively. The stationary distribution of the coupled system is uniform. Moreover, since both $Q_X$ and $Q_Y$ are normal, their Kronecker sum $Q$ is also normal. For simplicity, the observable is chosen to be the total number of phosphorylated sites among the three sites, i.e.
\begin{equation}\label{eq:bioObservable}
f(X_t,Y_t)=n_P(X_t)+Y_t,
\end{equation}
where $n_P(U) = 0$, $n_P(S) = n_P(T) = 1$, and $n_P(ST) = 2$.

For this system, the power spectrum is given by
\begin{equation}
G_f(\omega)=\frac{a(a^2+b^2+\omega^2)}
{(a^2+b^2+\omega^2)^2-4b^2\omega^2}
+\frac{r}{4r^2+\omega^2},
\end{equation}
where $a=k_++k_-$ and $b=k_+-k_-$. A simple sufficient condition for a nonmonotonic $G_f(\omega)$ with positive $\eta_f$ is
\begin{equation}\label{eq:bioPeakCondition}
\frac{a(3b^2-a^2)}{(a^2+b^2)^3}
>\frac{1}{16r^3}.
\end{equation}
Clearly, when $a^2<3b^2$ and when the switching rate $r$ is sufficiently large, Eq. \eqref{eq:bioPeakCondition} is satisfied and hence the system is oscillatory. Fig.~2(b) compares our universal and normal bounds (Eqs.~\eqref{eq:universalthermal} and \eqref{eq:normalthermal}) with Dechant's two bounds (Eq.~\eqref{eq:dechant}) as $r$ changes. For large $r$, Dechant's estimates yield a zero lower bound, whereas our universal and normal bounds remain strictly positive.

\begin{figure}[htb!]
\centering\includegraphics[width=0.43\textwidth]{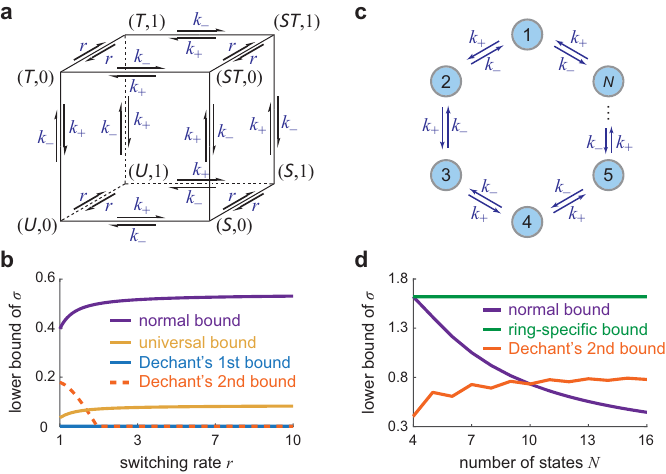}
\caption{\textbf{Thermodynamic bounds on the oscillation efficiency in Markov networks.}
\textbf{(a)} An idealized three-site phosphorylation model inspired by KaiC. Here $U$ denotes the state where both canonical sites are unphosphorylated, $S$ and $T$ denote the states where only one site is phosphorylated, and $ST$ denotes the state where both sites are phosphorylated; in addition, $0$ and $1$ denote the phosphorylation state of the additional site.
\textbf{(b)} Comparison of our universal and normal bounds with Dechant's two bounds for the model in (a).
\textbf{(c)} A uniform bidirectional ring.
\textbf{(d)} Comparison of our normal and ring-specific bounds with Dechant's second bound for the model in (c). In both (b) and (d), the parameters are chosen as $k_+ = 0.95, k_- = 0.05$.}
\label{figure2}
\end{figure}

To further compare our bounds with previous ones, we consider uniform bidirectional rings (Fig.~\ref{figure2}(c)). For simplicity, the observable is still chosen to be the first harmonic $f_i=\cos(2\pi i/N)$. For this observable, Dechant's two bounds in Eq.~\eqref{eq:dechant} can be computed explicitly as \cite{supp}
\begin{equation}\label{eq:Dring}
L_D^{(1)}
=\frac{\lambda_s}{c_N}\frac{\eta_f}{1-\eta_f},
\qquad
L_D^{(2)}
=\frac{\lambda_s}{c_N}\Phi(\eta_f),
\end{equation}
for any $\eta_f>0$, where
\begin{equation}
c_N:=\frac{(\Delta f)^2}{2}=
\begin{cases}
2,&N\text{ even},\\
(1+\cos(\pi/N))^2/2,&N\text{ odd}.
\end{cases}
\label{eq:ringRangeComparison}
\end{equation}
Comparing Eq.~\eqref{eq:Dring} with Eqs.~\eqref{eq:universalthermal} and \eqref{eq:normalthermal}, we can see that Dechant's two bounds are exactly our universal and normal bounds with the prefactor $4/(N-2)$ replaced by $1/c_N$. We emphasize that for uniform rings, Eq.~\eqref{eq:ringquadratic} gives a stronger bound with prefactor $1/\cos^2(\pi/N)$.

Fig.~\ref{figure2}(d) compares our normal and ring-specific bounds with Dechant's second bound for uniform rings. It is clear that our ring-specific bound performs the best among all estimates. Relative to Dechant's second estimate, our normal estimate is stronger for $N\leq9$, coincides with it for $N=10$, and is weaker for $N\geq11$.

\heading{Conclusions}
Here we have established universal dynamical and thermodynamic bounds on the oscillation efficiency in Markov networks. Unlike modal coherence, a high oscillation efficiency can benefit from non-normality and does not require complex eigenvalues. The thermodynamic bounds reveal the tradeoff between oscillation robustness, relaxation speed, and dissipation. Stronger results for normal networks and uniform rings further reveal how dynamical structure constrains oscillatory performance. These findings establish physical limits on the strength and cost of stochastic oscillations.

\heading{Acknowledgements}
The author acknowledges support from NSFC grant No. 12271020.

\footnotesize\setlength{\bibsep}{5pt}
\bibliography{myref}
\bibliographystyle{apsrev4-2}

\normalsize

\section*{End Matter}
\heading{Proof of Eq.~\eqref{eq:sectorN}}
For $i\ne j$, let $w_{ij}=\pi_iq_{ij}$ be the probability flow from state $i$ to $j$. Stationarity implies
\begin{equation}
\sum_{j\ne i}w_{ij}=\sum_{j\ne i}w_{ji},
\end{equation}
so the total stationary flow entering each state equals the flow leaving it. The cycle decomposition theorem claims that such a flow can be decomposed into positive flows around simple directed cycles \cite{qian1979decomposition, landim2016metastability}. Indeed, one can successively extract a directed cycle, subtract the smallest flow along its edges, and repeat. Each subtraction removes at least one positive-flow edge, so this procedure terminates after finitely many steps. Accordingly, we may decompose the operators $A$, $H$, and $K$ as
\begin{equation}
A=\sum_c A_c,\qquad
H=\sum_c H_c,\qquad
K=\sum_c K_c,
\end{equation}
where $c$ ranges over all directed cycles, $H_c=(A_c+A_c^*)/2$, and $K_c=(A_c-A_c^*)/2$. Moreover, we define the cycle energy $\EE_c(u)=\ip{u}{H_cu}$, so that $\EE(u)=\sum_c\EE_c(u)$.

Let $c=(x_0,\ldots,x_{\ell-1},x_0)$ denote a directed cycle of length $\ell\geq 2$, and let $a_c>0$ denote the common stationary flow around the cycle. For any observables $u,v\in\HH$, we write $u_j=u(x_j)$ and $v_j=v(x_j)$. The contributions of the cycle $c$ to the generator's bilinear forms are
\begin{equation}
\begin{split}
\ip{u}{A_cv}
&=a_c\sum_{j=0}^{\ell-1}
\overline{u_j}(v_j-v_{j+1}),\\
\ip{u}{H_cv}
&=\frac{a_c}{2}\sum_{j=0}^{\ell-1}
\overline{(u_{j+1}-u_j)}(v_{j+1}-v_j),\\
\ip{u}{K_cv}
&=\frac{a_c}{2}\sum_{j=0}^{\ell-1}
\overline{u_j}(v_{j-1}-v_{j+1}),
\end{split}
\end{equation}
with indices understood to be modulo $\ell$. In particular, the cycle energy is given by
\begin{equation}
\EE_c(u)=\ip{u}{H_cu}
=\frac{a_c}{2}\sum_{j=0}^{\ell-1}|u_{j+1}-u_j|^2.
\end{equation}
These expressions involve the same flow on every cycle edge, even when the stationary probabilities are unequal.

Define the unitary discrete Fourier transforms by
\begin{equation}
\widehat{u}_m
=\frac{1}{\sqrt{\ell}}\sum_{j=0}^{\ell-1}
u_j e^{-\ii\theta_m j},
\quad
\widehat{v}_m
=\frac{1}{\sqrt{\ell}}\sum_{j=0}^{\ell-1}
v_j e^{-\ii\theta_m j},
\end{equation}
where $\theta_m=2\pi m/\ell$. The inverse transform and Parseval identity give
\[
u_j=\frac{1}{\sqrt{\ell}}\sum_{m=0}^{\ell-1}
\widehat{u}_m e^{\ii\theta_m j},
\qquad
\sum_{j=0}^{\ell-1}\overline{u_j}v_j
=\sum_{m=0}^{\ell-1}\overline{\widehat{u}_m}\widehat{v}_m.
\]
Since shifting $v_j$ to $v_{j\pm1}$ multiplies its $m$th
Fourier coefficient by $e^{\pm\ii\theta_m}$, we obtain
\begin{equation}
\begin{split}
\ip{u}{H_cv}
&=\frac{a_c}{2}\sum_{m=0}^{\ell-1}
|e^{\ii\theta_m}-1|^2
\overline{\widehat{u}_m}\widehat{v}_m\\
&=a_c\sum_{m=0}^{\ell-1}
(1-\cos\theta_m)
\overline{\widehat{u}_m}\widehat{v}_m,\\
\ip{u}{K_cv}
&=\frac{a_c}{2}\sum_{m=0}^{\ell-1}
(e^{-\ii\theta_m}-e^{\ii\theta_m})
\overline{\widehat{u}_m}\widehat{v}_m\\
&=-\ii a_c\sum_{m=0}^{\ell-1}
\sin\theta_m\,
\overline{\widehat{u}_m}\widehat{v}_m.
\end{split}
\end{equation}
Consequently,
\begin{equation}
\EE_c(u)
=a_c\sum_{m=0}^{\ell-1}
(1-\cos\theta_m)|\widehat{u}_m|^2.
\end{equation}
These identities express the cycle bilinear forms as sums over independent Fourier modes.

The constant mode $m=0$ contributes to neither form, because
$1-\cos\theta_0=\sin\theta_0=0$. For every nonconstant mode,
$m=1,\ldots,\ell-1$, the half-angle identities imply
\begin{equation}
\frac{|\sin\theta_m|}{1-\cos\theta_m}
=\left|\cot\frac{\theta_m}{2}\right|
=\left|\cot\frac{\pi m}{\ell}\right|
\le \cot\frac{\pi}{\ell}.
\end{equation}
Indeed, $\pi m/\ell$ lies in $[\pi/\ell,\pi-\pi/\ell]$, and $|\cot x|$ attains its maximum on this interval at its endpoints. Combining this bound with the Cauchy-Schwarz inequality yields
\begin{equation*}
\begin{split}
&|\ip{u}{K_cv}|
\le a_c\sum_{m=1}^{\ell-1}
|\sin\theta_m|\,|\widehat{u}_m|\,|\widehat{v}_m|\\
&\le \cot\frac{\pi}{\ell}
\sum_{m=1}^{\ell-1}
\left[\sqrt{a_c(1-\cos\theta_m)}\,|\widehat{u}_m|\right]
\left[\sqrt{a_c(1-\cos\theta_m)}\,|\widehat{v}_m|\right]\\
&=\cot\frac{\pi}{\ell}
\sqrt{\EE_c(u)\EE_c(v)}.
\end{split}
\end{equation*}
Every simple cycle has length at most $N$. Summing over cycles and applying the Cauchy-Schwarz inequality once more, we obtain
\begin{align*}
|\ip{u}{Kv}|
&\le \sum_c|\ip{u}{K_cv}|\\
&\le \cot\frac{\pi}{N}
\sum_c\sqrt{\EE_c(u)\EE_c(v)}\\
&\le \cot\frac{\pi}{N}
\sqrt{\EE(u)\EE(v)}.
\end{align*}
Now substitute $u=H^{-1/2}x$ and $v=H^{-1/2}y$. Since
\begin{equation}
\EE(u)=\|x\|_\pi^2,\qquad
\EE(v)=\|y\|_\pi^2,
\end{equation}
the above inequality becomes
\begin{equation}
|\ip{x}{By}|
\le \cot\frac{\pi}{N}\|x\|_\pi\|y\|_\pi.
\end{equation}
Taking the supremum over unit vectors proves
\begin{equation}
\|B\|\le\cot\frac{\pi}{N}.
\end{equation}
This argument uses the cycle decomposition, rather than an eigenvector decomposition of the full generator, and therefore applies to nonnormal networks as well.

\heading{Sharpness for three and four states}
Here we show that for three-state and four-state networks, the dynamic bound in Eq.~\eqref{eq:dynamic} is sharp. We now give the explicit construction of families of three-state and four-state networks. They have one-way edges and do not satisfy thermodynamic consistency. Their efficiencies approach the universal bound as a singular parameter tends to zero.

For $N=3$, choose $0<\varepsilon<1/3$, and set
\begin{equation}
r=\frac{2\varepsilon}{1-\varepsilon},\quad s=1.
\end{equation}
The generator and centered observable are chosen to be
\begin{equation}\label{eq:Q3}
Q^{(3)}_\varepsilon=
\begin{pmatrix}-r&r&0\\0&-r&r\\s&0&-s\end{pmatrix},\qquad
f=(-\varepsilon,-\varepsilon,1-\varepsilon).
\end{equation}
The stationary distribution of the system is given by
\begin{equation}
\pi=\left(\frac{1-\varepsilon}{2},
\frac{1-\varepsilon}{2},\varepsilon\right).
\end{equation}
Using Eq.~\eqref{eq:psd}, it is easy to obtain the power spectrum
\begin{equation}
G^{(3)}_\varepsilon(\omega)
=\frac{2\varepsilon(\omega^2+3r^2)}
{\omega^4+(2r^2+s^2)\omega^2+r^2(r+2s)^2}.
\end{equation}
For $0<\varepsilon<1/3$, the power spectrum is nonmonotonic and attains its unique maximum at
\begin{equation}
\omega_*^2
=r(s-r)
=\frac{2\varepsilon(1-3\varepsilon)}
{(1-\varepsilon)^2}.
\end{equation}
The oscillation efficiency is then given by
\begin{equation}\label{eq:eta3}
\eta_\varepsilon^{(3)}
=\frac{(s-r)^2}{(r+2s)^2}
=\frac{(1-3\varepsilon)^2}{4}\longrightarrow\frac14,
\end{equation}
as $\epsilon\to0$. In particular, note that the two nonzero eigenvalues of $A=-Q$ are the roots of the quadratic equation
$z^2-(2r+s)z+r^2+2rs = 0$. When $\varepsilon<1/9$, the two nonzero eigenvalues are both real; however, the system has a positive oscillation efficiency. This reveals that complex eigenvalues are not necessary for a Markov network to have positive oscillation efficiency.

For $N = 4$, choose $0<\varepsilon<1/2$, and set
\begin{equation}
r=\frac{\varepsilon}{1-\varepsilon},\quad
s=1.
\end{equation}
The generator and centered observable are chosen to be
\begin{equation}
Q^{(4)}_\varepsilon
=
\begin{pmatrix}
-s & s  & 0  & 0\\
0  & -r & r  & 0\\
0  & 0  & -s & s\\
r  & 0  & 0  & -r
\end{pmatrix},
\qquad
f=(1,0,-1,0).
\end{equation}
The stationary distribution of the system is given by
\begin{equation}
\pi=\left(\frac\varepsilon2,\frac{1-\varepsilon}2,
\frac\varepsilon2,\frac{1-\varepsilon}2\right)
\end{equation}
and the power spectrum is given by
\begin{equation}
G^{(4)}_\varepsilon(\omega)
=\frac{2\varepsilon s(\omega^2+2r^2)}
{\omega^4+(r-s)^2\omega^2+4r^2s^2}.
\end{equation}
For $0<\varepsilon<1/2$, the power spectrum is nonmonotonic and attains its unique maximum at
\begin{equation}
\omega_*^2
=\sqrt{2}\,r(r+s)-2r^2
=\frac{\varepsilon(\sqrt{2}-2\varepsilon)}
{(1-\varepsilon)^2}.
\end{equation}
The oscillation efficiency is then given by
\begin{equation}
\eta_\varepsilon^{(4)}=
\frac12\left[1-(\sqrt2-1)
\frac{\varepsilon}{1-\varepsilon}\right]^2
\longrightarrow\frac12,\label{eq:eta4}
\end{equation}
as $\epsilon\to0$. Eqs.~\eqref{eq:eta3} and \eqref{eq:eta4}, together with Eq.~\eqref{eq:dynamic}, indicate that the dynamic bound in Eq.~\eqref{eq:dynamic} is sharp for three-state and four-state networks. We emphasize that here sharpness is an approached supremum, rather than an attained maximum.

\end{document}